\documentclass[12pt]{article}
\usepackage{graphics}

\title{Familon emission by dense magnetized plasma}

\author{N.V.~Mikheev\thanks{E-mail address:mikheev@univ.ac.ru},
E.N.~Narynskaya\thanks{E-mail address:elenan@univ.ac.ru}\\
{\small\it Division of Theoretical Physics, Department of Physics,}\\
{\small\it Yaroslavl State University, Sovietskaya 14,}\\
{\small\it 150000 Yaroslavl, Russian Federation.}}

\date{}

\begin{document}

\maketitle

\begin{abstract}

Emission of a familon caused by the processes $e^- \to e^- +  \phi$,
 $ e^- \to \mu^- +  \phi$ in dense magnetized plasma is investigated
in the model in which a familon
 have both direct and no direct coupling to
leptons via plasmon.
 The process probabilities and the integral familon action on plasma
 are calculated.
It is shown that the $ P$ odd interference phenomenon
in the process $e^- \to \mu^- + \phi$
 leads to the familon  force acting
on plasma along the magnetic field.

\end{abstract}

The familon, the
Nambu -- Goldstone boson, associated with the  spontaneous breakdown
of a global family symmetry
is of interest not only in theoretical aspect of elementary
particle physics, but in some astrophysical and cosmology 
applications~\cite{Raf}.
In particular, 
 through coupling to electron and
  photons,
familon could give a contribution to the energy and
momentum losses by stellar object. By this means, the
investigations of the familon involving processes under extreme
conditions, high matter densities and strong magnetic field, are
important  for  an analysis of some astrophysical cataclysms such as a
supernova explosion.

Here we study  forbidden in vacuum processes of  the familon
cyclotron emission $e^- \to e^-  +  \phi $, transition $e^- \to \mu^- +
\phi$ and their contributions into the energy losses by
 magnetized plasma. We consider the   physical situation
when
  the typical energy  of the plasma electrons, $E$,
is the largest physical parameter:
\footnote{ We use natural units in which $c=\hbar=1$, $e > 0$ is the
elementary charge. }
\begin{equation}
E^2 \gg eB \gg m^2_e.  \label{eq:cond1}
\end{equation}
The condition (\ref{eq:cond1}) corresponds to the relatively weak
magnetic field, when plasma electrons occupy highest
Landau levels. At the same time the magnetic field is still strong
enough in comparison with the  Schwinger value $ B \gg B_e, $ $ B_e
= m_e^2/e \simeq 4.41\times10^{13} $ G. Such extreme conditions: high
density of matter $ \sim 10^{14} g/cm^3 $, large electron  chemical
potential   $ \mu \sim 500 m_e,$ strong magnetic field up to $ B
\sim 10^{17} G$ could exist, for example, in the core of the
 exploding supernova. Notice, that in an external
magnetic field, the result of calculations depends not only
on typical kinematical invariants like $m^2$ and $p^2$, but also
on the field invariant
\begin{displaymath}
e^2(FF) = -2e^2B^2,
\end{displaymath}
and dynamical field invariant
\begin{displaymath}
e^2(pFFp) = e^2B^2E^2sin^2\theta,
\end{displaymath}
where
 $p_\mu$ is the particle 4-momentum,
$F_{\mu\nu}$ is the
tensor of the external magnetic field,
$\theta$ is the angle between the particle
 momentum $\vec p$ and the  magnetic field direction.
Inside the parentheses the tensor indices are contracted
systematically, for example $(pFFp) = p_\alpha 
F_{\alpha\beta}F_{\beta\nu}p_\nu.$

 Thus the conditions (\ref{eq:cond1}) can be
 rewritten in the invariant form:
\begin{displaymath}
e^2(pFFp) \gg [e^2(FF)]^{3/2} \gg m^6_e.
\end{displaymath}
It is known, that the problem of studies of the quantum processes
under the conditions (\ref{eq:cond1}) reduces to a  calculation in
the constant crossed field~\cite{Rit}. It is because that in  the
rest frame of a high energy electron, a  relatively weak
and smooth external electromagnetic field looks very close to the
constant crossed field $(\vec B \perp \vec E, \mid \vec B \mid =
\mid \vec E \mid)$, where $(FF) =  (\tilde F  F) = 0$,
$\tilde F^{\mu\nu}$ is the dual tensor of the external field. So,
the result  depends actually on the dimensionless dynamical
parameter $\chi$ only
\begin{equation}
\chi^2 =   \frac{e^2 (pFFp)}{m^6},
\label{eq:par}
\end{equation}
where $m$ is the mass  of  a particle.

\section {Familon cyclotron emission.}

\begin{figure}[b]
\centerline{\includegraphics{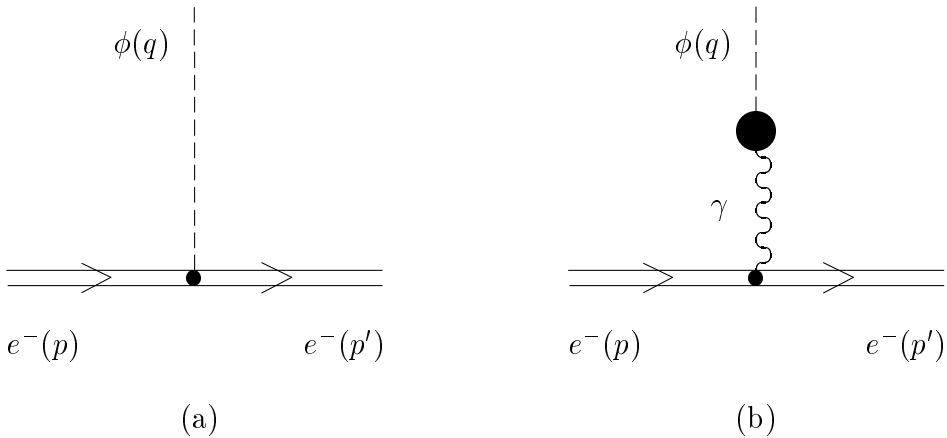}}
\caption{The Feinman diagrams  for the familon cyclotron emission
by plasma electron
in the presence of a magnetic field.}
\label{fig:eef}
\end{figure}

The  familon cyclotron emission by plasma
electron has two possible channels shown in  Fig.\ref{fig:eef}.
The process $ e^-  \to e^- + \phi  $ due to the direct familon --
electron coupling (Fig.\ref{fig:eef}a) can be described  by the
effective ${\cal L}$agrangian in the following form:
\begin{equation}
  L_{\phi e}= \frac{g_{\phi e}}{2m_e}
  [\overline \Psi_e \gamma_\alpha \gamma_5 \Psi_e] \partial_\alpha \Phi,
\label{eq:lag1}
\end{equation}
where $\Phi$ and $\Psi_e$ are the familon  and electron fields respectively,
$g_{ \phi e} = 2m_e/F$, $F$  is the family symmetry breaking scale.
The astrophysical constraint gives
$g_{\phi e} < 1,4\times10^{-13}$ ($F > 7\times10^9 GeV$)~\cite{PDG}.

\begin{figure}[t]
\centerline{\includegraphics{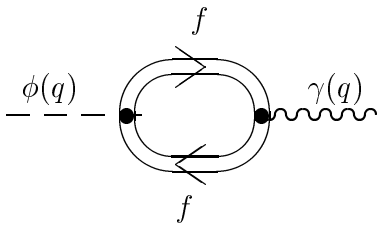}}
\caption{The effective  familon-photon interaction in an external magnetic
field}
\label{fig:loop}
\end{figure}

It is important that magnetic field induces a new effective interaction
between a familon and a photon of the type:
\begin{equation}
  L_{\phi \gamma}= g_{\phi\gamma} (\partial_\alpha A_\beta) 
\tilde F^{\alpha\beta} \Phi,
\label{eq:lag2}
\end{equation}
where $A_\mu$ is the four potential of the quantized
electromagnetic field, $g_{\phi\gamma}$ is the effective familon
-- photon coupling constant in the presence of external field,
which could be  derive from the diagram of Fig.\ref{fig:loop}, where double lines
indicate that the influence of the external field is taken into
account  in the propagators of virtual fermions $f$.
%
This constant $g_{\phi\gamma}$ could be extracted from the
paper~\cite{Mikheev} where the effective  field -- induced
interaction of a pseudoscalar particle with photons was
investigated. Notice that if the magnetic field is not so strong
($ B \ll B_\mu = m^2_\mu/e \simeq 1,8 \times 10^{18} G$, $m_\mu$
is the mass of the muon), only  the virtual electron gives a
contribution to the $g_{\phi\gamma}$, so for the $g_{\phi\gamma}$
coupling one can obtain the following result :
\begin{displaymath}
g_{\phi \gamma} = \frac{\alpha}{\pi}\, \frac{ g_{\phi e}}{m_e}.
\end{displaymath}
It should be stressed that the  familon -- photon interaction becomes possible
in the presence
of external magnetic field only. This is due  to the fact that
familon does not have both anomalous $\Phi G \tilde G$ and
 $\Phi F \tilde F$
coupling in vacuum ($G$ and $F$ are gluonic  and electromagnetic
fields, respectively). As a result, the emission of a familon via plasmon
intermediate state (Fig.\ref{fig:eef}b)
becomes possible in the presence  of both components of active medium:
plasma and magnetic field.

The $S$ -- matrix element of the process $e^- \to e^- + \phi$
 is:
\begin{displaymath}
  S = \frac{1}{F\sqrt{2\omega V}}
  \left [ \frac{2\alpha e}{ \pi } (q\tilde F G^L I^{(em)})
    - (q I^{(5)}) \right ],
\end{displaymath}
where $ I^{(em)}_\mu = \int d^4 x \overline \psi_e (p',x) \gamma_\mu  
\psi_e (p,x) e^{iqx}$,
 $I^{(5)}_\mu = \int d^4 x \overline \psi_e (p',x) \gamma_\mu \gamma_5 
\psi_e (p,x) e^{iqx}$,
 $\psi_e$ is the solution of the Dirac equation in the
constant crossed field~\cite{Ber}, $q^\alpha=(\omega,\vec q)$ is
the familon 4 - momentum, while $p^\alpha=(E,\vec p)$ and
$p'^\alpha=(E',\vec p')$ are the four-momenta of the initial and
final electrons,
 $ G^L_{\alpha\beta}$ is the longitudinal plasmon propagator.

At first glance, the amplitude of the familon emission via plasmon
contains the suppression associated with the fine structure
constant  $\alpha$. However, as the analysis show, the
contribution of both familon emission channels  into the process
could  be of the same order. The main motivation why  the  familon
emission via plasmon intermediate state could be expected to be
nonnegligible is that this channel has a resonant character
 at a
particular energy of the emitted familon.
It is  provided by the fact that
intermediate plasmon is longitudinal.
Similarly to the axion -- plasmon interplay~\cite{Mik}
the familon and the
longitudinal plasmon dispersion curves always cross for certain
energy $\omega = \omega_0$. At the same time the contribution  from the
transverse intermediate  plasmon turn out to be negligible small
in the case of ultrarelativistic plasma electrons. The  longitudinal
plasmon propagator $ G^L_{\alpha\beta}$ in the limit of weak
magnetic field (\ref{eq:cond1}) can be written as
\begin{displaymath}
  G^L_{\alpha\beta} \simeq  \,\frac{l_\alpha l_\beta}{q^2 -
  \Pi^L}, \,\,\,    l_\alpha = \sqrt{\frac{q^2}{(uq)^2 - q^2}} \left (
  u_\alpha - \frac{(uq)}{q^2}q_\alpha \right ).
\end{displaymath}
Here $l_\alpha$ and $\Pi^L$ are the eigenvector and eigenvalue of
the polarization operator corresponding to the longitudinal
plasmon, respectively,
 $u_\alpha$ is the four - vector of the  velocity of the medium.

In order to obtain  the probability  of the cyclotron familon emission
by a plasma electron we need to integrate over the phase space of
final particles  taking into account  the electron statistical factor.

The result of our calculations can be presented in the  medium
rest frame in the following form:
\begin{eqnarray}
 W_{(e^- \to e^- \phi)}  & \simeq &
\frac{1 }{2 \pi^2 F^2 E}
 \int_0^E   \frac{(E - \omega)d\omega}{e^{(\mu-E+\omega)/T} + 1}
  \times
  \label{eq:ver1}
 \\
 &  \bigg\lbrace & \frac{4\alpha^2(eB)^2}{3\pi }
   \frac{cos^2\theta}{\omega^2(1-\Pi^L/q^2)^2}
 +   3^{1/6} \Gamma  (2/3) m_e^2
  \left ( \frac{ \omega^2 eB sin\theta }{E^2(E - \omega)^2}\right )^{2/3}
  \bigg\rbrace,
  \nonumber
\end{eqnarray}
where $ \theta $ is the angle between the initial electron
 momentum $\vec p$  and the  magnetic field direction.

\section {Transition $e^- \to \mu^- +  \phi$.}

\begin{figure}[b]
\centerline{\includegraphics{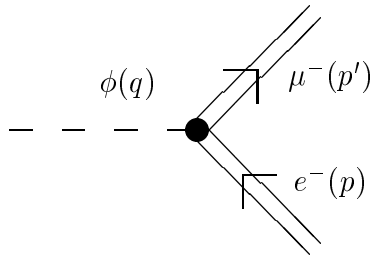}}
\caption{The Feinman diagram  for transition $e^- \to \mu^- + \phi$
in the presence of magnetized plasma.}
\label{fig:transition}
\end{figure}

The phenomenon of  fermion mixing gives rise to flavor --
nondiagonal
 familon -- fermion interaction. As a result, the processes with
 lepton number violation of the type $e^- \to \mu^- + \phi$,
 $\mu^- \to e^- + \phi$ becomes possible.
In this section we investigate the transition $e^- \to \mu^- + \phi$
as an additional channel of the familon emission by plasma electron
(Fig.\ref{fig:transition}).
This process was studied earlier~\cite{Averin} in the strong magnetic 
field limit under the condition
of smallness of the difference $(\mu - m_\mu)$,
  when $\mu^2 - m_\mu^2 \ll eB$,  so the final muons are produced in
the lowest Landau level only.
In contrast to~\cite{Averin} we consider as an example the conditions 
in a supernova
core with the relatively weak magnetic field,
 when $\mu^2 - m^2_\mu \gg eB$, and therefore  muons can occupy a
great number of excited Landau levels.

The interaction of
familon with electron and muon is described by effective
${\cal L}$agrangian in the form:
\begin{equation}
  L = \frac{1}{F}
  [\overline \Psi_\mu \gamma_\alpha (a + b\gamma_5)\Psi_e +
   \overline \Psi_e \gamma_\alpha (a + b\gamma_5)\Psi_\mu]
   \partial_\alpha \Phi,
 \label{eq:lag3}
\end{equation}
where
$\Psi_\mu $ is the muon field, $a^2 + b^2 =1$.

The $S$ -- matrix element of the process $e^- \to \mu^- + \phi$
 can be obtained immediately from the Lagrangian (\ref{eq:lag3})
by means of substitution of solutions of the Dirac equation in the
crossed field:
\begin{displaymath}
  S = \frac{-1}{\sqrt{2\omega V} F} \int
  d^4 x \overline \psi_\mu (p',x) \hat q (a + b\gamma_5)\psi_e (p,x)
  e^{iqx},
\end{displaymath}
where $p'^\alpha = (E',\vec p')$,  $p^\alpha = (E,\vec p)$
 are the muon and electron 4 -- momenta respectively.

Under  the conditions we consider the
 muon dynamical parameter $\chi$, $\chi^2 = e^2
(pFFp)/m^6_\mu$,
is rather
small and in this case the
result for the probability has a simple form:
\begin{equation}
  W_{( e^- \to \mu^- \phi)}=
\frac{m_\mu eB}{18 \sqrt 3  \pi F^2} \,sin\theta\, e^{-\sqrt 3 / \chi},
  \label{eq:ver2}
\end{equation}

At first  glance the probability (\ref{eq:ver2}) is
exponentially small in comparison with (\ref{eq:ver1}), however, as
it will be shown  below, the  plasma energy loss  via the familon
emission caused by cyclotron process $e^- \to e^- + \phi$ and
transition $e^- \to \mu^- + \phi$ turn out to be of the same order. This
is due to the fact that
 expression (\ref{eq:ver2}) in
contrast to (\ref{eq:ver1}) does not contain a suppression associated
mainly with the  smallness of electron mass $m_e$  or fine structure
constant $\alpha$.

\section {Familon emissivity.}
In the studies of the familon  involving processes, not only
probabilities but also the integral familon action on plasma is of
practical interest for astrophysics.
To illustrate possible
astrophysical applications of the result obtained
we estimate the energy losses of plasma via the
 familon emission:
\begin{equation}
  \dot \varepsilon =
       \frac{1}{(2\pi)^3}\int \frac{dW}{d\omega}\, \omega\, d\omega \,
       \frac{ d^3p}{e^{(E-\mu)/T}+1},
\label{eq:act}
\end{equation}
where $\dot \varepsilon$ is volume density of the plasma energy
losses per unit time, $\omega $ is the energy of the emitted
familon, $dW/d\omega$ is the differential probability of the process
considered.

The volume density of the
plasma energy losses
 caused by the familon emission can be presented as the sum
 of two contribution:
\begin{displaymath}
 \dot \varepsilon_{\phi} = \dot \varepsilon_{e^- \to e^- \phi}
  + \dot \varepsilon_{e^- \to \mu^- \phi}.
\end{displaymath}
Upon integrating over the phase space of initial electron and familon energy
in the case of  degenerate plasma, $\mu \gg T$, for the  cyclotron process 
contribution we find:
\begin{equation}
\dot \varepsilon_{e^- \to e^- \phi} \simeq \frac{1}{\pi^4 F^2}
 \left [ c_f m^2_e \left (
 \frac{eB}{\mu}\right )^{2/3} T^{13/3} +
  \frac{\alpha}{12}\frac{(eB)^2\omega_0^3}{(e^{\omega_0/T} -1)} \right ],
 \label{eq:eps}
\end{equation}
where
\begin{displaymath}
 c_f = \frac{14}{81} \left (   \frac{3}{4} \right )^{1/6}
  \Gamma^3 \left (\frac{1}{3} \right)  \zeta \left (   \frac{13}{3} \right ) 
\simeq    3,38,
\end{displaymath}
and $\omega_0$ is the energy  where the familon and longitudinal
plasmon dispersion curves cross~\cite{Mik}:
\begin{displaymath}
   \omega_0^2 \simeq  \frac{4\alpha}{\pi}\mu^2 \left ( ln\frac{2\mu}{m_e} -
      1\right ).
\end{displaymath}
The  first term in (\ref{eq:eps}) describes the contribution from
the process with  direct familon -- electron interaction
and the  second term defines the
familon emission via the longitudinal plasmon intermediate state.

For a contribution of the process $e^- \to \mu^- + \phi$
to the  plasma energy losses
 we obtain the following result:
\begin{equation}
 \dot \varepsilon_{e^- \to \mu^- \phi} \simeq 
 \frac{\sqrt{2} m^4_\mu \mu^3}{216
 \pi^{5/2} F^2}\,\frac{I(y)}{y^{3/2}},
   \label{eq:loss}
\end{equation}
\begin{displaymath}
 I(y) = \int^\infty_0 x^{7/2}
  \frac{e^{-y/x} dx }{e^{\frac{\mu}{T}(x-1)}+1},
 \,\,\,   y = \sqrt 3 \,\,\frac{m^3_\mu}{ eB\mu}.
\end{displaymath}

 The integral $I(y)$ in (\ref{eq:loss}) can be easily calculated in two
limiting cases:

i) the  case  of zero temperature limit
\begin{displaymath}
 \dot \varepsilon_{e^-  \to \mu^- \phi}
    \simeq \frac{\sqrt{2} m_\mu^4 \mu^3 }{216
 \pi^{5/2} F^2}\,\frac{e^{-y}}{y^{5/2}},
\end{displaymath}

ii)  relatively hot relativistic plasma ($T > \mu / y$)
\begin{displaymath}
 \dot \varepsilon_{e^-  \to \mu^- \phi}
 \simeq \frac{m^4_\mu \mu^3}{216\pi^2 F^2} \,\sqrt{2y}\,
   \left(  \frac{T}{\mu} \right )^{5/2}\,
 e^{-y(2-1/t)/t}, \,\,  t = \sqrt {\frac{y T}{\mu}}.
\end{displaymath}

We estimate  the familon emissivity under the conditions
which could be realized in a supernova explosion. As an example  we
take  $\mu = 250 MeV$, $T = 35 MeV$ and
$B = 10^{17} G$:
\begin{equation}
 \dot \varepsilon_{\phi} \simeq 10 ^{27}
 \left[  0.56_{ (e \to e \phi)} +  0.81_{ (e \to e \phi)^{(\gamma)}} +
        0.82_{ (e \to \mu \phi)} \right]
        \left(  \frac{7\times10^{9}GeV}{F} \right )^2
        \left ( \frac{erg}{cm^3 sec} \right).
\label{eq:res1}
\end{equation}

It is  seen from (\ref{eq:res1}) that the contributions
from process $e^- \to \mu^- + \phi$ and familon cyclotron emission
$e^- \to e^- + \phi$ are of the same order for the parameters
considered, while  the
 familon  luminosity $L_\phi \sim 10^{46} erg/sec $ is much less
 than the
neutrino luminosity $L_\nu \sim 10^{52} erg/sec $  from
supernova core during first few seconds after collapse.

It should be pointed that
at the cooling stage when the temperature becomes of order of MeV
the process $e^- \to \mu^- + \phi$ begins to  dominate over the cyclotron
emission $e^- \to e^-  + \phi$:
\begin{displaymath}
 \dot \varepsilon_{\phi} \simeq  \dot \varepsilon_{e^- \to \mu^- \phi}
  \simeq   0.92 \times 10 ^{25} \left ( \frac{7\times10^{9}GeV}{ F} \right)^2
  \left ( \frac{erg}{cm^3 sec}  \right)
\end{displaymath}
 and could provide a competition with the
neutrino energy losses at this stage, $\dot
\varepsilon_\nu \sim 10^{26}$ $  erg/cm^3 sec $  \cite{Raf}.

The another interesting feature of the processes considered
is the asymmetry of familon emission:
\begin{displaymath}
  A = \frac{1}{\dot \varepsilon} 
       \int \frac{q_3 \, d^3p\, /\,(2\pi)^3}{e^{(E-\mu)/T}+1}\,dW ,
\end{displaymath}
where $q_3$ is the  component (parallel
to field) of the  familon momentum.
Our calculations show that only
the transition $e^- \to \mu^- + \phi$
gives a contribution into this asymmetry
which has  very simple form:
\begin{equation}
  A \simeq \frac{ab}{3} \frac{eB}{m^2_\mu}.
\label{eq:as}
\end{equation}
\begin{figure}[h]
\centerline{\includegraphics{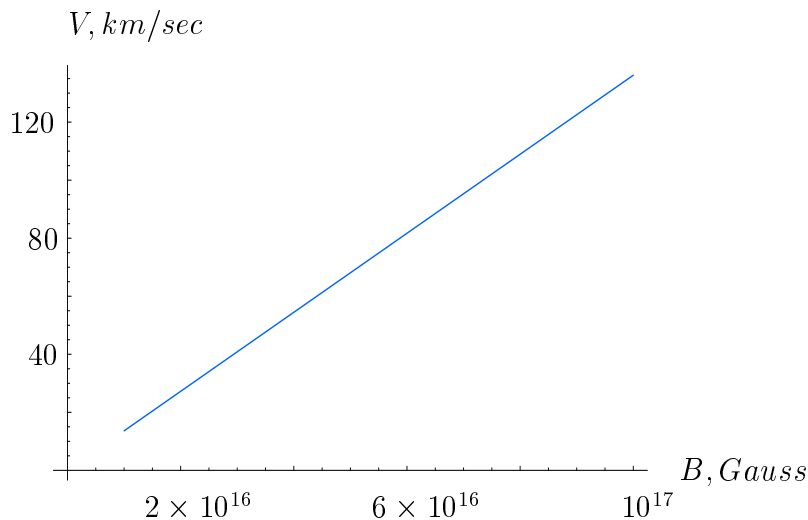}} \caption{The kick
velocity as a function of the  magnetic field strength.}
\label{fig:graf}
\end{figure}

We emphasize that asymmetry (\ref{eq:as})
 is caused by the
  interference  of the vector and axial -- vector couplings in the
  effective ${\cal L}$agrangian (\ref{eq:lag3}).

 As one can see, the dependence of $A$ on plasma parameters --
 the  electron chemical potential
 $\mu$ and temperature $T$,  totaly cancelled.
 Remind, however, the result (\ref{eq:as}) was obtained under assumption
 that $\mu$ is the largest physical  parameter of the task.

 The asymmetry of familon emission  leads to the familon force
 action on the magnetized plasma along  the magnetic field
 which in turn leads to the kick velocity of a supernova remnant.
 The result of our estimation of supernova remnant kick velocity
 is presented in Fig.\ref{fig:graf}.

  As one can see from
 Fig.\ref{fig:graf},  on the scale of the magnetic field
 up to $10^{17}$ Gauss the kick velocity does not exceed $100$ km/sec.
 So, the familon emission asymmetry can not solve the problem
 of proper  velocity of the pulsars.

In conclusion, we have studied the familon emission by plasma electrons
via the processes $e^- \to \mu^- + \phi$, $e^- \to e^- + \phi$. Due to
very weak interaction of familon with matter the above processes
could be important in astrophysics and cosmology. As a specific
application of the results obtained we have calculated the plasma
energy - momentum losses via the familon emission in a supernova
explosion.

\bigskip

This work was supported in part by the Russian Foundation
for Basic Research under the Grant No. 01-02-17334
and
 the Ministry of Education
of Russian Federation under the Grant No. E00-11.0-5.

\end{document}